\documentclass[final,5p,times,twocolumn,number,sort&compress]{elsarticle}

\usepackage{amssymb}
\usepackage{lipsum}
\usepackage{amsmath,amssymb}
\usepackage{physics}
\usepackage{graphicx}
\usepackage{dcolumn}
\usepackage{bm}
\usepackage{hyperref}
\usepackage[caption=false]{subfig}
\usepackage{subcaption}
\usepackage{multirow}
\usepackage[dvipsnames]{xcolor}
\usepackage{color}
\usepackage{cancel}
\usepackage[normalem]{ulem}
\usepackage{array}
\usepackage[utf8]{inputenc}
\usepackage{graphicx} 
\usepackage[T1]{fontenc}
\usepackage{xr}
\usepackage{float}
\usepackage{placeins}
\usepackage{makecell}
\usepackage{xr-hyper} 
\usepackage{hyperref}

\newcommand{\bea}{\begin{eqnarray}}
\newcommand{\eea}{\end{eqnarray}}

\DeclareUnicodeCharacter{2212}{-}

\newcommand{\BKA}[1]{\textcolor{red}{BKA: #1}}

\journal{Physics Letters B}

\begin{document}

\begin{frontmatter}


\title{Conflation of Ensemble-Learned Nuclear Mass Models for Enhanced Precision}
           
\author[first]{Srikrishna Agrawal}
\ead{agrawalsrikrishna0411@gmail.com}
\author[second,third]{ N. Chandnani}
\author[fourth]{T. Ghosh}
\author[third]{G. Saxena }
\ead{gauravphy@gmail.com}
\author[fifth]{B. K. Agrawal}
\author[fourth]{N. Paar}
\affiliation[first]{organization={Department of Mechanical Engineering, BITS-Pilani Hyderabad campus},
            city={Hyderabad},
            postcode={500078}, 
            country={India}}
 \affiliation[second]{organization={Department of Physics, School of Physical and Biological Sciences, Manipal University Jaipur},
            city={Jaipur},
            postcode={303007}, 
            country={India}}           
\affiliation[third]{organization={Department of Physics (H\&S), Govt. Women Engineering College},
            city={Ajmer},
            postcode={305002}, 
            country={India}}
\affiliation[fourth]{organization={Department of Physics, Faculty of Science, University of Zagreb},
            addressline={Bijeni\v cka cesta 32}, 
            city={Zagreb},
            postcode={10000}, 
            country={Croatia}}            
\affiliation[fifth]{organization={Saha Institute of Nuclear Physics},
            addressline={1/AF Bidhannagar}, 
            city={Kolkata},
            postcode={700064}, 
            country={India}}

\begin{abstract}

Ensemble learning algorithms, the gradient boosting and bagging regressors, are employed to correct the residuals of nuclear mass excess for a diverse set of six nuclear mass models. The weighted average of these corrected residuals reduces due to their partial cancellation, yielding a significant improvement in nuclear mass predictions. Our conflated  model, which integrates ensemble learning and model averaging (ELMA), achieves a root mean square error of approximately 65 keV, well below the critical threshold of 100 keV, for the complete data set of Atomic Mass Evaluation (AME2020). The validity of ELMA is demonstrated through the evaluation of $Q$ values for $\alpha$ decay, showing a marked decrease in deviations from experimental data relative to predictions from individual nuclear mass models. We have also compiled a table of nuclear mass excesses and binding energies for about 6,300 nuclei, which serves as a valuable resource for various nuclear physics applications and is publicly accessible via the ELMA web interface (\url{https://ddnp.in}).
\end{abstract}

\begin{keyword}
Machine learning  \sep  Nuclear masses \sep Binding Energies \sep Q-Values 


\end{keyword}

\end{frontmatter}




\section*{Introduction}
\label{introduction}
The mass of an atomic nucleus, or its mass excess, is one of the fundamental observables that encapsulates total binding energy, thereby serving as a direct insight to the underlying nuclear interactions and many-body quantum structure \cite{chartier2002mass}. Accurate values of mass are necessary for mapping the limits of nuclear existence at the proton and neutron drip lines \cite{chartier2002mass}. 
It also plays an important role in astrophysical reaction network calculations, determining nuclear abundances not only through rapid and slow neutron capture processes but also in other nucleosynthesis pathways. The  uncertainties in nuclear masses of few hundred keV can change calculated elemental abundances by many orders of magnitude \cite{arnould2007r, martin2016impact}. 
Despite several technological advances in mass measurements \cite{lunney2017extending, smith2008first, hausmann2001isochronous, yamaguchi2013challenge, czihaly2025exploring}, experimental determination is exceedingly challenging for nuclei far from line of stability, which are characterized by extremely short half-lives, low production rates \cite{chartier2002mass}, and the complicated task of separating low-energy isomeric states from the ground state \cite{giesel2024mass}. Therefore, one relies heavily on theoretical calculations for unexplored territory of the nuclear chart \cite{boehnlein2022colloquium}.

Atomic Mass Evaluation (AME-2020) provides a comprehensive and reliable database of experimentally measured masses for over 2400 nuclei \cite{huang2021ame}, however, theoretical predictions of masses of the nuclei away from the line of stability remain indispensable. Theoretical models predict the existence of over 6000 isotopes, spanning vast regions of the nuclear chart, particularly on the neutron-rich side, many of which are experimentally inaccessible \cite{audi2006history, wapstra2003ame}. A multitude of models have been developed, that is broadly divided into the macroscopic-microscopic approaches and fully microscopic theories. The earliest of these is the Liquid Drop Model, whose binding energy described by the foundational Bethe-Weizsäcker semi-empirical mass formula \cite{bethe1936nuclear, Bhagwat2014PRC90}, its predictive accuracy is limited, with root mean square error (RMSE) of about 3 MeV from experimental data \cite{huang2021ame}. Later macroscopic-microscopic refinements led to the Finite-Range Droplet Model (FRDM) which significantly led to reduction of RMSE to 0.662 MeV \cite{moller2016nuclear}. An alternative approach, the Duflo-Zuker (DZ) family of models, was developed based on a shell-model Hamiltonian, achieving RMSE of around 0.3–0.6 MeV \cite{duflo1995mass}. Lately, the Weizsäcker-Skyrme (WS4) model delivered one of the highest accuracies in this class with an RMSE of 0.298 MeV \cite{wang2014mass}. An improved version, the WS4+ model uses radial basis functions to correct systematic errors, obtaining a very small discrepancy of less than 0.250 MeV for most heavy nuclei \cite{ma2019nuclear}. Microscopic models are often based on self-consistent mean-field frameworks. Non-relativistic Hartree-Fock-Bogoliubov (HFB) theory became highly competitive with global models achieving RMSE below 0.8 MeV \cite{aboussir1995mass, goriely2013skyrme}. Relativistic approaches, framed within Covariant Density Functional Theory (CDFT), have also been extensively developed \cite{vretenar2005rhb, zhou2016mccdf}. Early Relativistic Mean-Field (RMF) models had RMSE of more than 1 MeV \cite{ring1996rmf}, though later versions with corrections have improved this to ~0.5 MeV \cite{zheng2014mass}. These global theoretical models generally achieve an accuracy in predicting experimental data in the range of RMSE 0.2–0.8 MeV. But their predictive performance deteriorates near nuclei far from the beta-stability line, where the deviations can increase to several MeV \cite{utama2017bayesian}. This uncertainty poses a major challenge towards consistent extrapolation of exotic nuclear properties and is well below the sub-100 keV precision needed to perform high-fidelity r-process calculations \cite{shang2024dnn}.

The accuracy gap of theoretical models created a ground for the application of Machine Learning (ML), which emerged as a powerful tool for improving nuclear mass predictions. The current state-of-the-art in ML for the prediction of nuclear mass is dominated by the use of a residual learning strategy, where algorithms are trained on the difference between experimental data and theoretical predictions for nuclear mass excess \cite{boehnlein2022colloquium}. A key demonstration of this approach was presented by Utama \textit{et al.}, who used a Bayesian Neural Network (BNN) to improve the predictions of several mass models by training it on their "residuals" \cite{utama2016nuclear}. The approach reduces the RMSE of the FRDM on the validation set from 0.664 MeV to 0.374 MeV. and, more importantly, offered a framework to compute the statistical uncertainty of the predictions \cite{utama2016nuclear}. Within this research paradigm, a diverse array of architectures has achieved high precision. Foundational kernel methods like Gaussian Process Regression (GPR) have demonstrated strong performance, achieving a RMSE as low as 0.26 MeV on test sets and an RMSE of 0.67 MeV when extrapolating to 71 newly measured nuclei in AME2020, confirming its predictive power on new data \cite{yuksel2024nuclear}. A significant improvement was achieved with a more advanced BNN model, which reached an accuracy of 84 keV \cite{niu2022nuclear}. Other advanced neural networks, such as a physics-informed Fully Connected Neural Network, have also shown further reduction with an RMSE of 0.122 MeV on its test set and extrapolation capability with an RMSE of 0.191 MeV for 71 newly measured nuclei in AME2020 \cite{huang2025validation}. Tree-based ensemble methods have also advanced the frontier of raw predictive accuracy; systematic studies applying algorithms like CatBoost to refine theoretical models have yielded RMSE of approximately 0.160 MeV in test data \cite{liu2025model}. More recently, a multi-task Gaussian Process model that simultaneously predicts mass and charge radii achieved a RMSE of 0.090 MeV in complete dataset for nuclear masses, highlighting the continued improvement in kernel-based methods \cite{Ye2025}.   

\par The pursuit of predictive accuracy is often hampered by model uncertainty, the ambiguity relies on selecting a single best model from multiple plausible candidates \cite{saito2024uncertainty,kejzlar2023local}. Standard statistical practice, which typically relies on a single "best" model, often ignores this uncertainty, leading to over-confident inferences and conclusions that may be less reliable than they appear \cite{alhassan2024bayesian}. This challenge is underscored by the No Free Lunch theorem, which posits that no single algorithm is optimal for every problem \cite{adam2019no}. To address this, model averaging has emerged as a powerful statistical tool, often implemented through what are known as ensemble methods \cite{kejzlar2023local}. The foundational principle of this approach is that a committee of diverse models, when combined with their predictions, can achieve greater accuracy and robustness than any individual member \cite{alhassan2024bayesian}. 
By aggregating the output of various models, where each captures different facets of data, ensemble techniques can effectively reduce overall error, mitigate the effects of individual bias of models, and improve generalization for new data \cite{pang2024studying}. 
This implementation is presented by Bentley \textit{et al.}, which introduces the Four Model Tree Ensemble \cite{bentley2025further}. The FMTE is the composite model created by taking weighted average of four distinct Least Squares Boosted Ensemble of Trees models. This multi-level averaging approach results in a model with an accuracy of RMSE 76 keV when evaluated against the AME 2020 experimental data. Furthermore, the model's robust extrapolation capability validated against post predicted mass measurements after AME2020 compilation, which achieved an RMSE of 0.376 MeV lower than the individual mass models considered \cite{bentley2025further}. 

In the present work, we employ diverse set of six nuclear mass models which can be broadly characterized as macroscopic-microscopic model, non-relativistic and relativistic self consistent mean-field models. The RMSE for the residuals of nuclear masses for these models for full AME2020 are in the range of 0.25-1.0 MeV. The raw residuals obtained from individual nuclear mass models are corrected using gradient boosting regressor (GBR) and bootstrap aggregation (BAR) \cite{friedman2001greedy, buhlmann2012bagging}, which are the most prominent ensemble learning methods \cite{liu2025model, Gao2021}. The GBR builds models sequentially, where each model is trained to correct the errors of its predecessors resulting reduced bias \cite{emami2023sequential, rizkallah2025enhancing}. The BAR trains multiple models on different random subsets of the data and averages their predictions to reduce variance, minimize overfitting, and improve reliability \cite{sutton2005classification, baskin2017bagging}. We obtain a single improved nuclear mass model by performing the weighted average over all the corrected residuals. Our conflated nuclear mass model obtained by integrating ensemble learning and model averaging yields an overall RMSE $\sim$ 65keV which is well below the critical value of 100keV. We also confront our mass model with the newly measured mass excess and for the $Q$-values of $\alpha$-decay. We have compiled values of nuclear masses and binding energies for about 6000 nuclei from our improved nuclear mass-model.  


\section*{Ensemble Learning}
Our predictive framework integrates two complementary ensemble techniques : Gradient Boosting Regressor (GBR) and Bootstrap Aggregating, or Bagging (BAR). Both models are well suited for capturing the complex, non-linear dependencies present in nuclear mass excess data, yet they address different aspects of model performance. The GBR focuses on sequentially reducing bias through iterative error correction \cite{friedman2001greedy, emami2023sequential}, while BAR targets variance reduction by aggregating predictions from multiple independently trained models \cite{buhlmann2002analyzing, buhlmann2012bagging}.  

\begin{figure}[t]
    \includegraphics[width=0.5\textwidth, keepaspectratio]{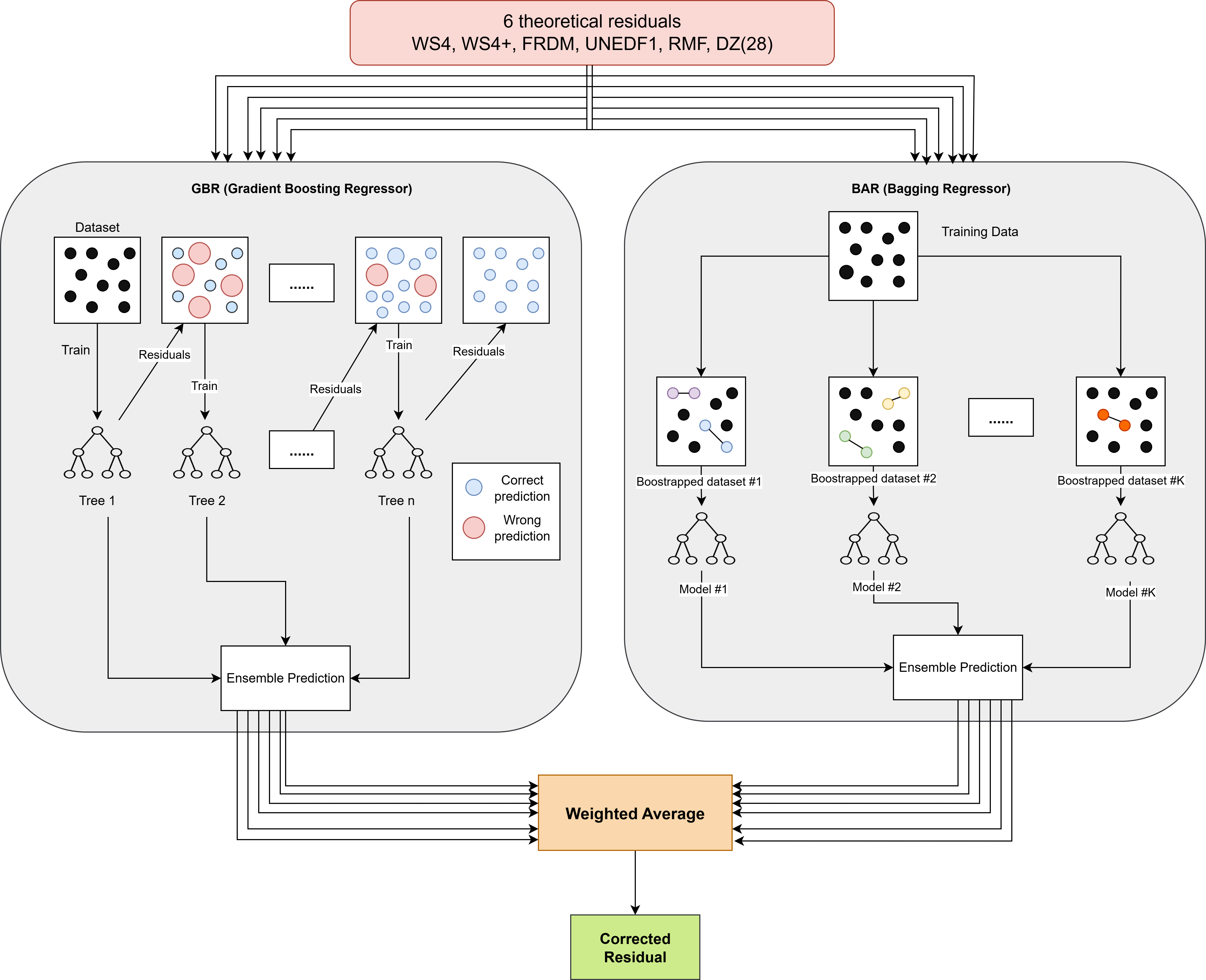}
    \caption{Workflow of GBR and BAR as implemented in this study to correct the raw residuals. GBR uses 5000 sequential estimators trained on residuals, BAR uses 25 parallel SVR estimators trained on bootstrap samples. The outputs are combined via weighted averaging.}
    \label{flowchart}
\end{figure}

Figure~\ref{flowchart} shows the pipeline implemented for both methods and the final prediction fusion step. In GBR (left panel), the model begins with the complete dataset for training and builds a sequence of 5000 decision tree estimators, each trained to correct the residual errors of its predecessors. The ensemble prediction is obtained by combining all estimators, weighted by the learning rate. In BAR (right panel), the dataset is resampled (25 times) with replacement to produce distinct bootstrap datasets. We use 
Support Vector Regressor (SVR) as a base estimator in the BAR algorithm for each of the datasets independently, and the outputs are averaged to produce the final prediction.

The outputs obtained from both GBR and BAR combined together, yielded a hybrid model balance the bias reduction properties of boosting with the variance-reduction properties of bagging. Complete details of hyperparameter optimization, and model configuration are provided in the Supplementary Material (SM).

\section*{Results and Discussion}
We have selected six nuclear mass models such as Weizsäcker-Skyrme (WS4) \cite{wang2014mass},  WS4+ \cite{ma2019nuclear},  finite-range droplet model (FRDM) \cite{moller2016nuclear}, Duflo Zuker model (DZ28) \cite{duflo1995mass}, Universal Nuclear Energy Density Functional 1 (UNEDF1) \cite{kortelainen2012nuclear} and Relativistic Mean-Field (RMF) \cite{Geng2003PTP, Geng2005Thesis} that represent a wide spectrum of theoretical approaches. The WS4 and its improved version WS4+ are macroscopic-microscopic models based on the Skyrme energy-density functional, with WS4+ including further empirical corrections for enhanced accuracy. The FRDM combines macroscopic  finite range droplet energy and microscopic shell and pairing corrections, providing reliable predictions for  masses across the nuclear chart. The DZ(28) is a parameter-rich Duflo–Zuker model that blends macroscopic and microscopic features for improved mass reproduction. The UNEDF1 is a self-consistent non-relativistic mean-field model based on an effective Skyrme energy-density functional for global modeling of nuclear properties. The variant of RMF model considered corresponds to interaction of nucleon through mesons along with the inclusion of nonlinear self- and mixed-interactions term and coupling strengths being density independent. These models collectively cover both empirical and fully microscopic theoretical frameworks, providing a broad and robust basis for our present analysis.

To improve the accuracy of nuclear mass prediction, the raw residuals (RR) are corrected using various machine learning algorithms.
The RR is defined as:
\bea
\delta M_i^{\rm RR} = M_{\rm Exp} - M_i \label{eq:delMRR}
\eea
where $M_{\rm Exp}$ is the experimental nuclear mass excess for a given
nucleus and $M_i$ is the corresponding value for the
$i$-th nuclear mass model. The corrected residuals (CR) are:

\bea
\delta M_i^{\rm CR} &=& \delta M_i^{\rm RR} - \delta M_i^{\rm ML} \label{eq:delMCR} \\
&=& M_{\rm Exp} - M_i^{\rm cor}
\eea

where the corrected mass excess $M_i^{\rm cor} = M_i + \delta M_i^{\rm ML}$,
with $\delta M_i^{\rm ML}$ being the machine-learned value of the raw residuals. We employed ensemble learning methods, namely the GBR and BAR, to estimate the values of $\delta M_i^{\rm ML}$ in Eq. (\ref{eq:delMCR}). The weighted average of residuals can be obtained as:

\begin{align}
\delta \bar{M}^{\rm X} &= \sum_{i=1}^{N_X} w_i^X \, \delta M_i^{\rm X} 
\label{eq:delMX_BAR}
\end{align}
where {\rm X} is either RR or CR. $N_{RR}$ = $N_{m}$ and $N_{CR}$ = 2$N_{m}$, with $N_{m}$ the number of mass-models considered. The normalized weights, $w_i^X$ is given by,

\bea  
w_i^X = \frac{\left( \frac{1}{\sigma_i^{\rm X}} \right)^2}{\sum_{k=1}^{N_X} \left( \frac{1}{\sigma_k^{\rm X}} \right)^2}
\eea
with $\sigma_i^{\rm X}$ is the root mean square error for a given nuclear mass model with raw or corrected residuals and can be evaluated as:

\bea
\sigma_i^{\rm X} = \sqrt{\frac{1}{N_d} \sum_{j=1}^{N_d} \left( \delta M_i^{\rm X}(j) \right)^2}
\eea
where the index $j$ runs over all nuclei of a given dataset of
size $N_d$. Thus, the models that perform better or having smaller values of $\sigma_i^{\rm X}$ are given higher weights, contributing  more effectively to improving the accuracy of the prediction. These weights are calculated using the RMSE values from test dataset.

The models were trained using 12 features, as detailed in Table~2 of the SM, with the target variable being $\delta M_i^{\rm RR}$ as given by Eq. (\ref{eq:delMRR}).
The training dataset is constructed from the Atomic Mass Evaluation  (AME2020) by excluding the nuclear mass excess values of 71 nuclei measured for the first time, as described in the SM. After training, the model’s performance for the corrected residuals is evaluated on various datasets-labeled by test, extpl, 2020, new, and full as described  in Table~3 of the SM. 

 \begin{table}[H]\vspace{-0.5em} 
\centering
\footnotesize
\setlength{\tabcolsep}{4pt}
{\renewcommand{\arraystretch}{1.1} 
\begin{tabular}{c|l|cccccc}
\hline
&  & \multicolumn{6}{c}{$\boldsymbol{\sigma}$ (MeV)} \\
\cline{3-8}
& $N_{\text{m}}$ & \textbf{train} & \textbf{test} & \textbf{extpl} & \textbf{2020} & \textbf{new} & \textbf{full} \\
\hline
\multirow{3}{*}{\rotatebox{90}{$\overline{\mathbf{RR}}$}} 
& 2  & 0.1886 & 0.1874 & 0.3263 & 0.1938 & 0.2893 & 0.1953 \\
& 4  & 0.1996 & 0.1978 & 0.3299 & 0.2043 & 0.2813 & 0.2054 \\
& 6  & 0.2004 & 0.1976 & 0.3285 & 0.2049 & 0.2775 & 0.2059 \\
\hline
\multirow{4}{*}{\rotatebox{90}{$\overline{\mathbf{CR}}$}} 
& 2 & 0.0484 & 0.1186 & 0.2013 & 0.0718 & 0.2627 & 0.0771 \\
& 4 & 0.0426 & 0.1195 & 0.1784 & 0.0671 & 0.2311 & 0.0715 \\
&6& 0.0412 & 0.1181 & 0.1702 & 0.0654 & 0.2223 & 0.0695 \\
\cline{2-8}
& MAE &{0.0299} & {0.0835} & {0.1247} & {0.0404} & {0.1565} & {0.0419} \\
\hline
\end{tabular}}
\caption{Root mean square errors $\sigma$  for the weighted-average model obtained using raw residuals (RR) and corrected residuals (CR) for different numbers of nuclear mass models ($N_m$), obtained using Eq. \ref{eq:delMX_BAR}. Corrected residuals are obtained by using ensemble learning algorithm, GBR and BAR. Results are presented for six datasets—train, test, extpl, 2020, new, and full dataset. The values of mean absolute errors (MAE) for CR case with $N_m = 6$ are also presented in the last row.}
\label{tab1}
\end{table}\vspace{-0.8em} 

\begin{figure*}[!htbp]
    \centering
    \includegraphics[width=0.97\textwidth, keepaspectratio]{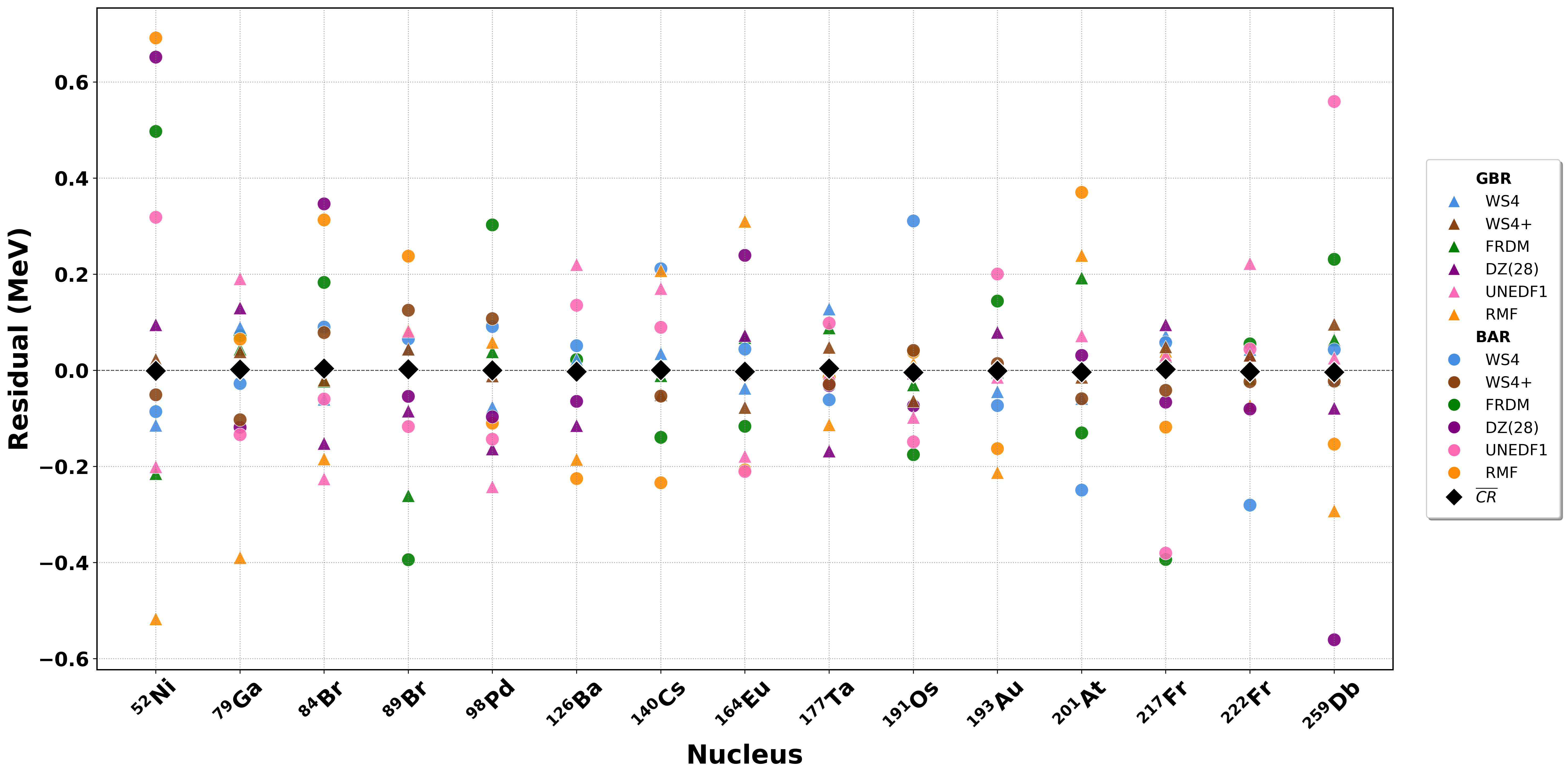}
\caption{Comparison of the weighted average nuclear mass excess values, obtained using corrected residuals, with those predicted individually by each nuclear mass model employing GBR and BAR methods for a set of selected nuclei. The magnitude of weighted average CR are considerably smaller than those for the individual models, primarily due to partial cancellation of model-specific deviations.}
    \label{delM_bar_selected}
\end{figure*}

The weighted-average nuclear mass excess is calculated using Eq. (\ref{delM_bar_selected}) for $N_m = 2, 4,$ and $6$, considering both the raw and corrected residuals. The $N_m = 2$ case uses the WS4 and WS4+ models; for $N_m = 4$, residuals from FRDM and DZ(28) are further  included; and for $N_m = 6$, the average is taken over all six models considered. The RMSE values for WS4 and WS4+ are smaller than those of the other models (see Table~4 of SM). 
Table \ref{tab1} presents the RMSE values ($\sigma$) obtained using the weighted average of residuals for $N_m = 2, 4,$ and 6. For $N_m = 6$, the mean absolute error of the corrected residuals are  also listed in the last row. Each nuclear mass model yields two sets of corrected residuals, corresponding to the GBR and BAR ensemble approaches.
For RR, the values of $\sigma$ increase slightly with $N_m$ for most datasets, except for the 'new' dataset, which contains nuclei measured after AME2020. This behavior arises because the weights are heavily skewed: WS4 and WS4+ together account for nearly 82\% of the total weight, leaving only \textasciitilde{}18\% distributed among the other models (Table~4, SM). In contrast, for CR, the values of $\sigma$ generally decrease with increasing $N_m$. Here, the weights are more evenly distributed, with WS4 and WS4+ contributing about 60\% and the remaining models collectively contributing  ~40\%. This redistribution demonstrates how the GBR and BAR ensembles effectively integrate contributions from a wider range of models, mitigating the dominance of any single one. For the 'extpl' and 'new' datasets, $\sigma$ decreases by 20–25\% as $N_m$ increases from 2 to 6. To further substantiate this observation, we analyzed all 15 possible combinations of four models out of the six mass models considered and evaluated the corresponding RMSE reductions for both the extpl and new datasets. In every case, the conflated model achieved RMSE values lower by 30–70 keV compared to the best-performing individual model. This demonstrates that the improvement is not an artifact of choosing a particular subset of models but rather a robust statistical feature of the averaging approach. This marked reduction—especially for nuclei far from the $\beta$-stability line, underscores the superior predictive capability of the CR method compared to RR. Further, across all datasets (except 'new'), the $\sigma$ values from CR are reduced by more than 40\%; for the 'new' dataset, the reduction is about 20\%. 

\begin{figure}[H]
    \centering
    \includegraphics[width=0.80\columnwidth, height=0.50\textheight]{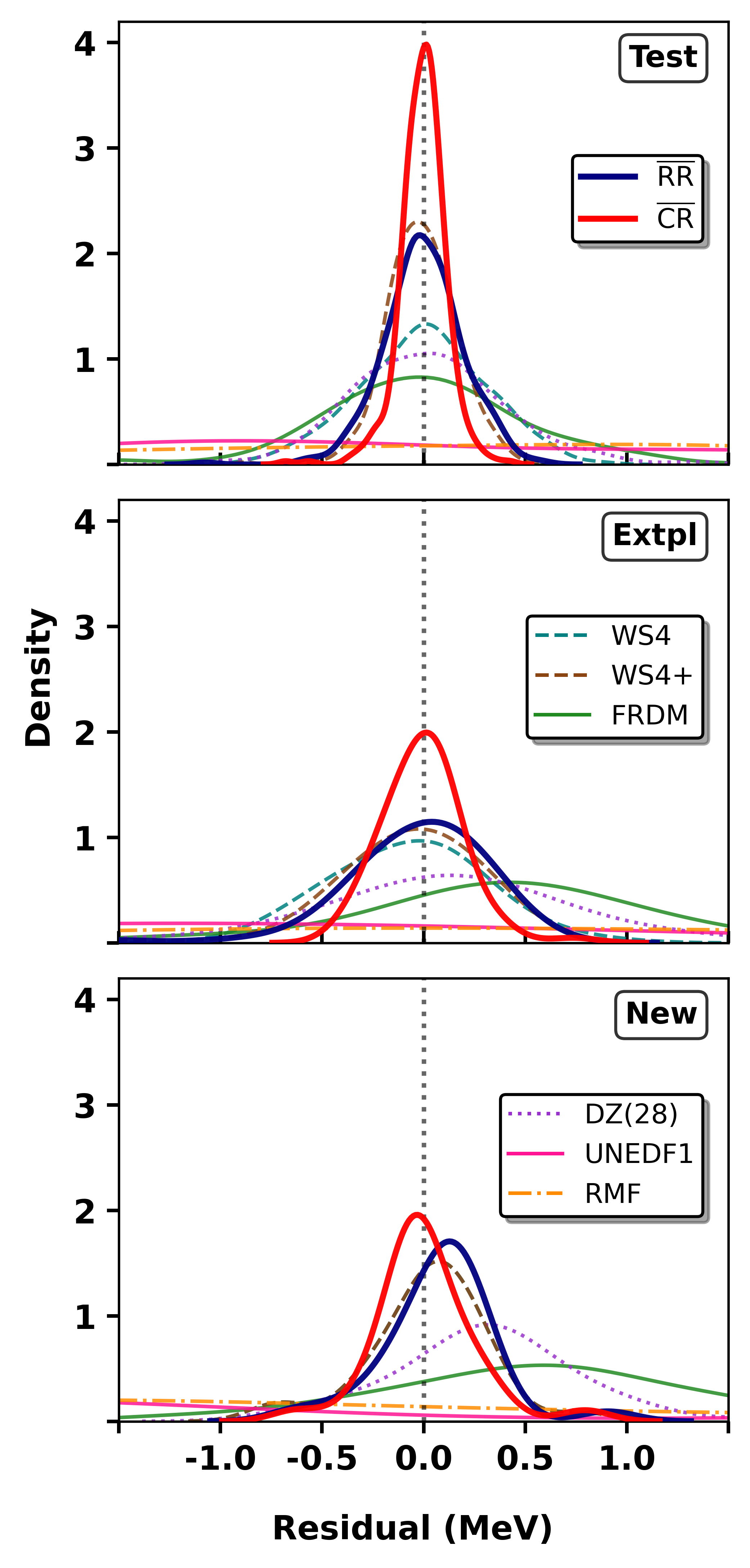}
    \caption{
    The probability density distributions for the weighted average of raw residuals (RR) and corrected residuals (CR) for the nuclear mass excess corresponding to test, extrapolation, and new datasets. The results for the RR from each of the six individual nuclear mass models are also plotted for comparison. The vertical dotted line indicates complete agreement between predictions and experimental data.}
    \label{resi_dist}
\end{figure}

Figure \ref{resi_dist} shows the probability density distributions for the weighted averages of both raw and corrected residuals, incorporating contributions from all six nuclear mass models. The upper panel corresponds to the weighted-average residuals for the 358 nuclei in the test dataset, the middle panel for the 71 nuclei in the extpl dataset, and the lower panel for the 31 nuclei in the new dataset measured after AME2020. For the comparison, we have also plotted the  distributions corresponding to the raw residual for the different nuclear mass models considered. For the raw residuals, the weighted-average distributions closely resemble those of WS4+, which accounts for more than 60\% of the total weight. In contrast, the distributions for the corrected residuals are more sharply peaked around zero for all the different datasets, owing to the redistribution of weights that allows contributions from the other mass models as well (see Tables~3 and 4 of SM). In the case of RR, the residual distributions for individual models are generally broad, which is associated with the higher values of $\sigma$, in particular for UNEDF1 and RMF. The peak heights of the probability density distribution corresponding to the test and extrapolated datasets obtained from the corrected residuals are nearly twice those of the raw residuals. For the dataset labeled 'new', however, this increase is only marginal. This is also evident from Table~\ref{tab1} that the overall performance is better for the weighted average values of nuclear masses in comparison to the corrected residuals for the individual model across the test, extrapolation, and newly measured data sets.

\begin{figure}[h!]
    \centering
    \includegraphics[width=0.98\columnwidth, keepaspectratio]{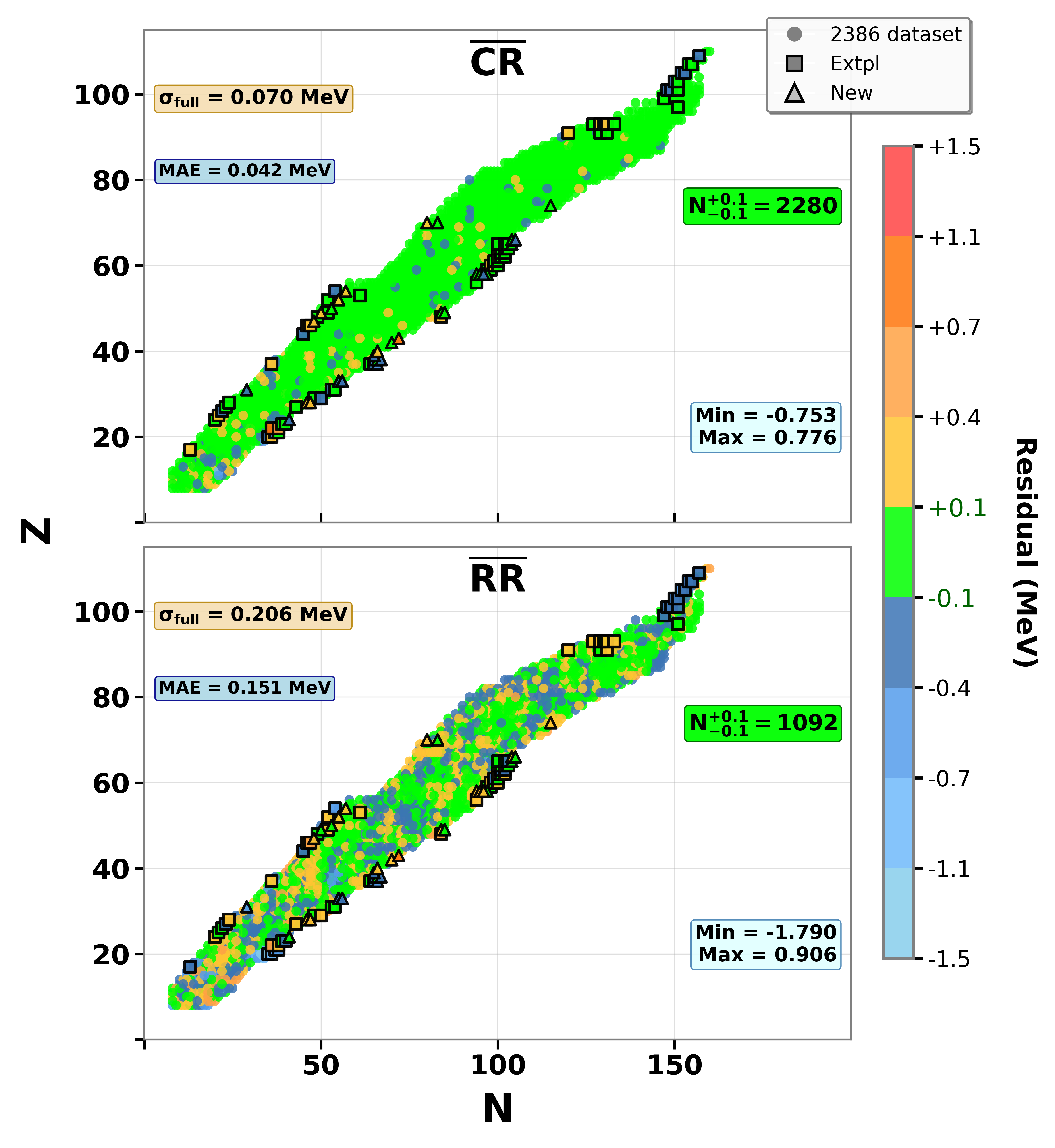}
    \caption{The values of weighted-average nuclear mass excess in the N-Z plane. The lower and upper panels show the results obtained using raw (RR) and corrected (CR) residuals, respectively. The $N^{+0.1}_{-0.1}$ denotes the number of nuclei whose predicted nuclear mass excess has an absolute error of less than 100 keV (green symbols).}
    \label{delM_av}
\end{figure}

Figure \ref{delM_av} shows the weighted average values of the raw residuals, $\delta\bar{M}^{\rm RR}$(lower panel), and the corrected residuals, $\delta\bar{M}^{\rm CR}$ (upper panel), for all nuclei in the complete data set. The green colour symbols indicates  the nuclei whose absolute value of residuals lie within 100 keV, total number of such nuclei is denoted by $N^{+0.1}_{-0.1}$. The value of $N^{+0.1}_{-0.1}$ increases significantly for the case of corrected residuals. In the data set comprising 2,386 nuclei, most have turned green in the upper panel, indicating improved agreement as compared to the case of RR. Several nuclei from the extrapolated (extpl) data set as seen in the upper panel have also changed to green as compared with the lower panel. The values of the $\sigma_{full}$= 0.070 MeV and the mean absolute error (MAE) = 0.042 MeV for the complete set obtained with weighted average corrected residuals are nearly three times smaller than those for the raw residuals. The maximum spread in the values of $\delta\bar{M}^{\rm CR}$ is almost 1 MeV smaller than $\delta\bar{M}^{\rm RR}$.

The advantage of the weighted-averaging approach in improving nuclear mass-excess predictions is evident from Table~\ref{tab1} and Figures \ref{resi_dist} and \ref{delM_av}. These effects are  further illustrated in Figure \ref{delM_bar_selected}, which compares the weighted-average of corrected residuals for a representative set of 15 nuclei, selected over a wide mass range, with those from individual nuclear mass models obtained using the GBR and BAR ensemble methods. It is evident that the weighted-average residual values lie very close to zero,   demonstrating  the cancellation effect and importance of the averaging procedure. 
Residuals, by definition, exhibit both positive and negative deviations and, under unbiased model assumptions, follow a distribution centered at zero with finite variance \cite{montgomery2021introduction}. This statistical symmetry implies that averaging residuals leads to partial cancellation of deviations, thereby reducing the net error. Consistent with this, the corrected residuals from individual models are randomly scattered around zero, often with opposite signs for the GBR and BAR approaches within the same model. It may also be noted that out of 12 different values of corrected residuals, for each of the 15 nuclei considered, we have roughly equal number of points on either side of zero line leading to partial cancellation of the residuals while performing the weighed average. This partial cancellation brings the weighted-average residuals, shown as black diamonds, consistently closer to zero than those of any single model. Models that overestimate the mass excess ($\delta\bar{M}^{\rm CR}$<0) for certain isotopes are balanced by others that underestimate them ($\delta\bar{M}^{\rm CR}$>0), reducing systematic bias. Because weights are assigned according to each model’s predictive accuracy on the test set, better-performing models exert greater influence on the final estimate, enhancing the stability of ensemble predictions. This mutual error balancing not only lowers RMSE values (Table~\ref{tab1}) but also yields more robust and reliable predictions across the nuclear chart, including extrapolative and data-sparse regions where individual models typically underperform as seen in Table~\ref{tab1} and Figure \ref{delM_bar_selected}.

\subsection*{Application of refined nuclear masses}

We name our refined nuclear mass model as ELMA which denotes
ensemble learning and model averaging. The refinement of nuclear mass predictions through the use of ELMA may provide a foundation for broader applications in nuclear physics and astrophysics. Having minimized the deviations and enhanced predictability across the nuclear chart, the resulting mass table becomes more reliable resource for theoretical and experimental studies alike. In nuclear structure and decay studies, accurate nuclear masses serves as a cornerstone for calculating separation energies, decay energies and reaction thresholds. Even small improvements in mass prediction accuracy can lead to substantial gains in the reliability of theoretical models and experimental planning.

One of the most direct and critical applications of an improved mass table is in the determination of the $Q_\alpha$ values. These values are required for determining the properties of alpha decay, assessing the boundaries of nuclear stability and interpreting decay chains of newly synthesized heavy and superheavy elements. For superheavy nuclei in particular, where experimental mass measurements are scarce or absent, the precision of predicted values of $Q_\alpha$ is often considered as deciding factor in confirming the discovery of new isotope. 

\begin{figure}[H]
    \centering
    \includegraphics[width=\columnwidth, height=0.35\textheight]{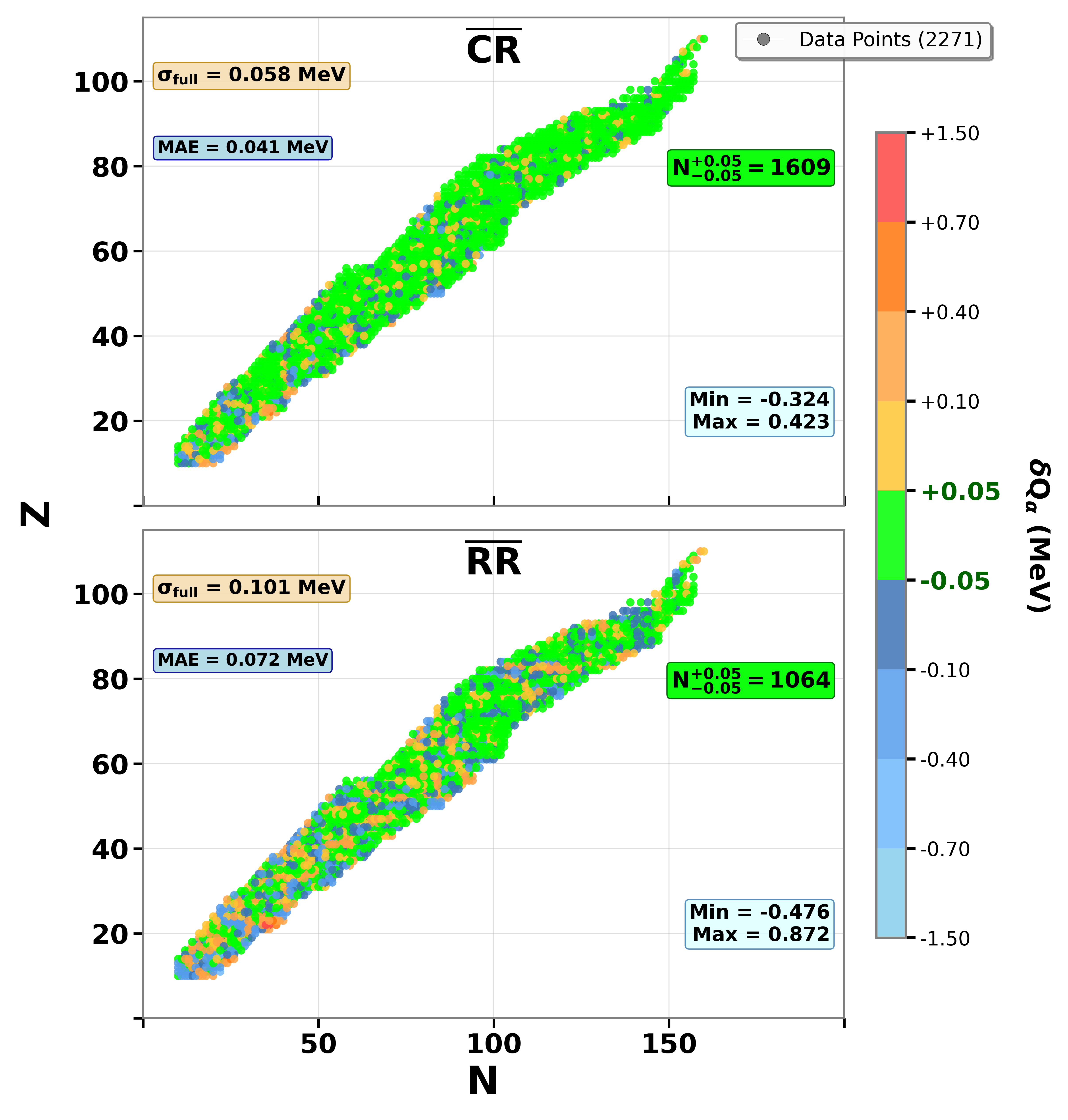}
    \caption{Residuals for $\alpha$-decay $Q$ values obtained using the weighted-average values of raw residuals (lower panel) and corrected residuals (upper panel) of the nuclear mass excess. The values of $\sigma$ and $N_{-0.05}^{+0.05}$ are also given. $N_{-0.05}^{+0.05}$ indicates the number of cases corresponding to $\left| \delta Q\alpha \right| < 50$ keV (green symbols).}
    \label{Q_alpha}
\end{figure}

In Figure~\ref{Q_alpha}  we plot residuals of the $Q_\alpha$ values  calculated with weighted averaged values of RR and CR for nuclear mass excess. The refinement in nuclear mass prediction is evident, since, the residual distribution of $Q_\alpha$ is narrower as can be visualized from figure by enhancement in the number of green symbols (upper panel). The RMSE reduces to 0.058 MeV (upper panel) which is factor of two smaller than those for the RR (lower panel). In addition, the number of nuclei for which  $\left\lvert \delta Q_\alpha \right\rvert < 0.05$
MeV increases from 1064 to 1609, reflecting a substantial enhancement in predictive accuracy of the $Q_\alpha$ values across a wide range of periodic chart. This enhancement in predictive accuracy is particularly critical, as an uncertainty of just $\pm$0.05 MeV in $Q_{\alpha}$ demonstrates a profound sensitivity of decay rates, where an increase of that magnitude may result in a predicted half-life changed by $\sim40\%$ of its original value.

\subsection*{Nuclear Mass Table for ELMA}
The values of mass excess and binding energies obtained using nuclear mass model ELMA are calculated for 6303 nuclei. This mass table is made available as supplementary material (\texttt{masstable\_ELMA.txt}). By providing, more accurate mass predictions, this table can be used to integrate into variety of research workflows, including r-process nucleosynthesis, $\beta$ decay energy calculations, and the evaluation of reaction Q-values in astrophysical environments. Furthermore, its enhanced precision in $Q_\alpha$ estimation makes it as valuable method for guiding experimental results in the synthesis and identification of isotopes at the limits of nuclear chart. We have also deployed ELMA on a web interface, accessible via the provided link (\url{https://ddnp.in}.
 In the future, we plan to continuously update ELMA by incorporating new experimental measurements as well as theoretical advancements, thereby enhancing its predictive accuracy.

\section*{Conclusions}

We employed ensemble learning and model averaging to improve the predictability  of nuclear mass models. Our study uses a diverse set of six models: WS4, WS4+, FRDM, DZ(28), UNEDF1, and RMF. The raw residuals for each of these nuclear mass models were employed to train two ensemble learning algorithms: GBR and BAR. The GBR iteratively reduces bias through sequential error correction, whereas the BAR reduces
variance by aggregating predictions from multiple independently trained models. The ensemble learned residuals were then used to correct the raw
residuals, thereby incorporating effects not captured by the original nuclear mass models. Hence, there are two different types of corrected residuals for each of the raw residuals for a given nuclear mass model. An improved mass model was obtained by conflating all twelve sets of corrected residuals through a weighted average.

 Our nuclear mass model, constructed from a weighted average of corrected residuals, demonstrates a substantial enhancement in predictive accuracy compared with the individual mass models across the validation datasets considered. The resulting conflated mass model achieves an RMSE of approximately 0.0654 MeV for the AME2020 dataset.
 It is found that the weights assigned to the corrected residuals are substantially redistributed compared to those for the raw residuals. Consequently, the nuclear mass predictions obtained from the weighted average of the corrected residuals exhibit a marked improvement, primarily due to partial cancellation of the residuals. This improvement becomes more pronounced with the increases in the number of nuclear mass models included in the averaging procedure  (see Table \ref{tab1}).

We have applied our mass model to evaluate the $Q$ values for $\alpha$ decay and compared them with experimental data. The deviations from the measured values are found to be significantly reduced. 
We designate our mass model as ELMA (Ensemble Learning and Model Averaging). Using ELMA, we have compiled a table of nuclear mass excess and binding energies for approximately 6,300 nuclei. The ELMA has been deployed on a web interface, accessible via the following URL: \url{https://ddnp.in}.

The authors gratefully acknowledge Jatin Fulwani for his valuable assistance in developing the ELMA web interface. 
B.K.A. acknowledges with gratitude the support received under the Raja Ramanna Chair scheme of the Department of Atomic Energy (DAE), Government of India.
T.G. and N.P. acknowledge support from the Croatian Science Foundation under the project Relativistic Nuclear Many-Body Theory in the Multimessenger Observation Era (HRZZ-IP-2022-10-7773). N.C. and G.S. expresses gratitude for the support received from the Department of Science \& Technology (DST), Government of Rajasthan, Jaipur, India under F24(1)DST/R\&D/2024-EAC-00378-6549873/819. 

\bibliographystyle{elsarticle-num}

\bibliography{references}





\end{document}



\twocolumn[
\begin{center}
\textbf{Supplementary Material}\\
\textbf{Conflation of Ensemble-Learned Nuclear Mass Models for Enhanced Precision}\\
Srikrishna Agrawal, N. Chandnani, T. Ghosh, G. Saxena, and B. K. Agrawal, N. Paar
\end{center}
\hrule
\vspace{1em}
]
In this supplementary material (SM), we present an overview of the ML models employed in our calculations: Gradient Boosting Regressor (GBR) and Bootstrap Aggregation (BAR). We describe different datasets employed for training and validating our machine learning models and provide details about the features considered. We also provide some important information about the raw residuals for the six different mass models considered and the corresponding corrected residuals. 
\section{Machine Learning Models}
\label{ML_models}


\subsection{Gradient Boosting Regressor}
Gradient Boosting Regressor (GBR) is an ensemble machine learning technique, highly effective for regression tasks involving complex and non-linear data relationships. GBR based upon the principle of boosting, wherein multiple predictive models (typically decision trees) are trained sequentially where each subsequent model aiming to correct the errors made by its predecessors \cite{friedman2001greedy}. 

GBR core concept involves fitting a series of regressors to the residuals (errors) of earlier models rather than to the target values. At each step, a new regressor is trained to predict the negative gradient of the loss function for the current model's predictions and to attempt in a stage-wise manner to minimize this loss function \cite{emami2023sequential}. This iterative approach creates a strong predictive ensemble by improving predictive accuracy of the model. 

A distinguish feature of GBR is its flexibility by using regression as a mean squared error in optimizing arbitrary differentiable loss functions. The learning process starts with an initial model, often a simple mean prediction, and iteratively adds new models that reduce the remaining error. The contribution of each new estimator is scaled by a learning rate parameter, which controls the magnitude of each incremental update. A learning which is smaller leads to better generalization, though it requires more iterations for boosting.

Several hyperparameters are critical to the GBR performance:
\begin{itemize}
\item \textbf{\texttt{n\_estimators}}: It specifies the boosting stages number (i.e., trees) to fit. Increasing this can typically improves accuracy to a point, but can also risk overfitting. 
\item \textbf{\texttt{learning\_rate}}: It controls the step size at each iteration. Lower values require more estimators, but help in smoothing the learning process, which in turn leads to superior performance.
\item \textbf{\texttt{max\_depth}}: Sets the maximum depth of each individual tree, thus controlling the complexity of the base learners. Shallow tress (often stumps with depth 1-3) are commonly used to prevent overfitting.
\item \textbf{\texttt{subsample}}: This represents the fraction of training samples used for fitting each base learner, introducing randomness and potentially enhancing robustness of the model.
\item \textbf{\texttt{random\_state}}: This ensures reproducibility by fixing the sequence of random operations during training.
\end{itemize}
In our study, we used the GBR class from the scikit-learn library, we took the n\_estimators as 5000, learning\_rate as 0.1, max\_depth taken as 3, subsample fixed to 0.8 and random\_state as 42, to balance the model expressiveness and complexity. Proper hyperparameter is essential to maximize generalization performance while minimizing the risk of overfitting, thereby ensuring that GBR provides both accurate and reliable predictions on unseen data.

\subsection{Bagging Aggregated Regressor}
Bootstrap Aggregating, or Bagging (BAR), is an ensemble learning technique that is primarily used to reduce the variance of predictive models by averaging outputs from multiple base estimators trained on different bootstrap samples of the data \cite{buhlmann2002analyzing}. In BAR, the dataset is divided into multiple subsets through a process known as bootstrapping, and each subset is used to train a separate model, often using the same machine learning algorithm, each model learns from a slightly different perspective because the subset contains different samples. Once the models are trained, they are used to make predictions on new, unseen data. The predictions from these individual models are then aggregated, typically by averaging in regression tasks, to produce a more robust final prediction \cite{buhlmann2012bagging}.

In our implementation, we used Support Vector Regressor (SVR) as the base estimator due to its capacity to model complex non-linear relationships. However, since SVR in high-dimensional spaces can be sensitive to data fluctuations and noise, combining it with bagging helps mitigate overfitting and stabilizes its performance across varied regions of data.

The BAR was constructed using the \texttt{Bagging Regressor} class from scikit-learn. Several hyper-parameters were tuned and are discussed below:
\begin{itemize}
    \item \textbf{\texttt{estimator}}: This specifies the base model that can be used in each bootstrap sample. In our case, it was an SVR pipeline that includes feature scaling and the SVR model itself. Using a strong base learner like SVR ensures that each bagged model is capable of learning complex relationships.
    \item \textbf{\texttt{n\_estimators}}: This is defined as the number of individual estimators (or SVR models) to be trained on different bootstrap samples. We experimented with multiple values, 5, 10, 25, and 50 to observe the relationship between predictive accuracy and computational complexity. A higher number generally leads to more stable and accurate predictions as the averaging effect becomes stronger. However, beyond a certain point, performance improvement diminishes, and computational cost increases significantly. Therefore, we set \texttt{n\_estimators=25}, as it offered the optimal trade-off between accuracy and computational cost. 
    \item \textbf{\texttt{max\_samples}}: This determines the fraction or number of training samples to be drawn (with replacement) for training each base model. It controls how diverse the individual models are. By using different subsets of data, the ensemble learns to generalize better and reduces the likelihood of overfitting to any single subset's noise. In our implementation, we set \texttt{max\_samples} = 0.6, which provided sufficient diversity among the base models while maintaining overall predictive stability.
    \item \textbf{\texttt{random\_state}}: A fixed random seed was used to ensure reproducibility across multiple runs. To ensure reproducibility across multiple runs and enable consistent comparison of results under different hyperparameter optimization, we fixed the \texttt{random\_state} to 42. 
\end{itemize}
By carefully configuring these parameters, BAR was able to significantly improve the model's resilience to overfitting while enhancing its performance in extrapolative regions \cite{dietterich2000ensemble}.
\subsection{Support Vector Regression}

The Support Vector Regression (SVR) \cite{drucker1996support} is a machine learning algorithm particularly suited for regression tasks, adapted from the foundational principles of Support Vector Machines (SVMs) originally designed for classification \cite{boser1992training}. Unlike ordinary least squares regression, which minimizes the squared error for all data points, SVR introduces a tolerance margin $\epsilon$ within which errors are not penalized. The model seeks a loss function that approximates the target values as closely as possible while ensuring that most data points fall within this $\epsilon$ insensitive zone \cite{geron2017hands}.


The defining aspect of SVR is its use of support vectors, data points that lie outside the $\epsilon$ margin, plays a crucial role in shaping the final regression function. The SVR algorithm attempts to construct a function that deviates from the true targets by no more than $\epsilon$, which helps in maintaining the flatness of the model. This trade-off is governed by a regularization parameter $C$, which balances the complexity of the model with its tolerance for training error \cite{geron2017hands}. 
The strength of SVR is seen when dealing with non-linear dependence of data on features through the use of the kernel trick. Kernels implicitly map the original input features into a higher-dimensional space where a linear relationship can be approximated. This transformation enables SVR to model complex patterns without explicitly computing the non-linear mapping. Among the available kernels, the commonly preferred kernel is the Radial Basis Function (RBF) kernel for its ability to capture localized data variations. The RBF kernel is defined as:


\begin{equation}
\begin{aligned}
K_G(x, x') &= \exp\left(-\gamma ||x - x'||^2\right).
\end{aligned}
\label{kernel}
\end{equation}

where, $x$ and $x'$ represent two input data points, and their Euclidean distance is denoted as $||x - x'||$, while $\gamma$ is the kernel coefficient. Eq. \ref{kernel} quantifies the similarity or dissimilarity between these data points, based on their distance in the input feature space. This results in higher similarity if data points are closer and conversely, lower similarity in data points that are more distant \cite{geron2017hands}. The $\gamma$ determines the influence of each data point. A lower value of $\gamma$ allows the model to consider points farther apart, resulting in smoother, more generalized decision boundaries. In contrast, a high $\gamma$ value restricts the influence to nearby points, enabling the model to capture more complex patterns but increasing the risk of overfitting. Careful tuning of $\gamma$ is particularly important when using non-linear kernels such as the RBF, as it directly affects the flexibility and expressiveness of the regression function.

The regularization hyperparameter, denoted as $C$, is essential in balancing between maximizing the margin and minimizing the training error. A smaller value of $C$ encourages a wider margin at the cost of more training errors, promoting generalization, whereas a larger $C$ places more emphasizes on accurately fitting of training data, potentially leading to overfitting.  To identify optimal values of the hyperparameter, we performed an intensive grid search by varying $C$ between 100 to 2000 and $\gamma$ from 0.05 to 1.0. This tuning was performed independently for each theoretical model. The $\epsilon =0.1$ is selected to define the width of the no-penalty zone, effectively allowing minor prediction errors without contributing to model loss. The values of optimized hyperparameter are presented in Table~\ref{tab:svr_hyperparameters}

\begin{table}[H]
\centering
\caption{SVR hyperparameters — kernel coefficient $\gamma$ and regularization parameter $C$, used during training for each nuclear mass model.}
\scriptsize 
\setlength{\tabcolsep}{8pt} 
\renewcommand{\arraystretch}{1.2} 
\begin{tabular}{lcccccc}
\hline
 & WS4 & WS4+ & FRDM & DZ(28) & UNEDF1 & RMF \\
\hline
$\gamma$ & 0.09 & 0.69 & 0.40 & 0.61 & 0.53 & 0.42 \\
$C$      & 167  & 611  & 609  & 1351 & 639  & 1569 \\
\hline
\end{tabular}
\label{tab:svr_hyperparameters}
\end{table}



\section{Features and Dataset}
\begin{table*}[htbp]
\centering
\small
\caption{List of 12 nuclear features used for learning the nuclear masses (see the text for details).}
\begin{tabular}{|c|l|c|p{10cm}|}
\hline
\textbf{\#} & \textbf{Feature} & \textbf{Symbol} & \textbf{Description / Formula} \\
\hline
1  & Proton Number          & $Z$                 & Number of protons in the nucleus. \\
2  & Neutron Number         & $N$                 & Number of neutrons in the nucleus. \\
3  & Mass Number            & $A$                 & Total number of nucleons: $A = Z + N$. \\
4  & Surface term      & $A^{2/3}$           & Surface area term: $A^{2/3}$. \\
5  & Isospin Asymmetry      & $I$                 & $I = \frac{N - Z}{A}$. \\
6  & Even-Odd Proton Indicator   & $Z_{\mathrm{eo}}$   & $Z_{\mathrm{eo}} = 0$ if $Z$ is even, $Z_{\mathrm{eo}} = 1$ if $Z$ is odd. \\
7  & Even-Odd Neutron Indicator  & $N_{\mathrm{eo}}$   & $N_{\mathrm{eo}} = 0$ if $N$ is even, $N_{\mathrm{eo}} = 1$ if $N$ is odd. \\
8  & Proton Magic Gap       & $v_{Z}$             & Distance from the nearest proton magic number: $v_Z = | Z - Z_{\text{magic}} |$. \\
9  & Neutron Magic Gap      & $v_{N}$             & Distance from the nearest neutron magic number: $v_N = | N - N_{\text{magic}} |$. \\
10 & Promiscuity Factor     & $PF$                & $PF = \frac{v_Z \cdot v_N}{v_Z + v_N}$. \\
11 & Proton Shell Number    & $Z_{\text{shell}}$  & Shell orbital index for protons (0, 1, 2, 3, 4 based on range). \\
12 & Neutron Shell Number   & $N_{\text{shell}}$  & Shell orbital index for neutrons (0, 1, 2, 3, 4 based on range). \\
\hline
\end{tabular}
\label{supplementary_tab1}
\end{table*}

A robust foundation for any machine learning model lies in both the selection of relevant input features and the careful partitioning of the dataset for training, testing and extrapolative evaluation. We employ a set of twelve physics informed features as described in (Table \ref{supplementary_tab1}), chosen to capture essential characteristic of nuclear structure and empirical regularities known to influence nuclear mass excess. These features were chosen based on their relevance to structure of nucleus and their proven efficacy in previous mass modeling studies \cite{yuksel2024nuclear}. The fundamental features include the proton number ($Z$) and neutron number ($N$), which specify the nuclide and incorporate well-established shell closures intrinsic to nuclear stability. The mass number ($A$) and its transformations, such as $A^{2/3}$, encode macroscopic size and surface effects consistent with liquid drop-inspired mass formulas.
The isospin asymmetry ($I = (N - Z)/A$) measures the difference between the number of neutrons and protons in a nucleus, and it plays an important role in the symmetry energy part of the nuclear equation of state. To account for pairing effects, which cause sudden changes in mass between nuclei with odd or even numbers of protons and neutrons (known as odd-even mass staggering), we include binary variables ($Z_{eo}$, $N_{eo}$) indicating whether the proton and neutron numbers are odd or even. These help the models capture sharp local variations in nuclear mass caused by nucleon pairing.

The magic number gap features ($v_Z$, $v_N$) characterize shell effects and are central to explaining localized maxima and minima of nuclear mass surfaces, quantifying how close a nucleus is to canonical magic numbers such as 8, 20, 28, 50, 82, 126, and 184. These features enable the model to implicitly identify shell closures and the associated enhanced stability. In addition, the promiscuity factor (PF), defined as $PF = \frac{v_Z v_N}{v_Z + v_N}$, captures the interplay between valence protons and neutrons and serves as a indicator for proton–neutron interaction strength. This feature helps describe collective effects beyond simple shell closures. Finally, the proton and neutron shell numbers ($Z_{\text{shell}}, N_{\text{shell}}$) provide discrete indices representing the shell model orbitals of the last proton and neutron. These indices (0–4) map the nucleon numbers into established shell ranges, such as 1–28, 29–50, 51–82, 83–126, and above 127, thereby encoding large-scale structural organization of the nuclear chart. 

The associated derived shell correction, pairing gaps and local nuclear structure anomalies supplement this feature content to give the model a multi-scale view of the mass differences.  Such a comprehensive set of physics informed features enables the models to learn complex patterns, including shell closures and structural irregularities that are often missed by purely empirical predictors.

\begin{table}[htbp]
\centering
\scriptsize
\caption{Various datasets required for training and validation of our model performance (see the text for details).}
\resizebox{\linewidth}{!}{%
\begin{tabular}{|c|c|c|c|c|c|c|}
\hline
\textbf{Set} & train & test & extpl & 2020 & new & full \\
\hline
\textbf{Number} & 2028 & 358 & 71 & 2457 & 31 & 2488 \\
\hline
\end{tabular}%
}
\label{supplementary_tab2}
\end{table}

We employ nuclear mass data from the most recent AME2020 compilation, which reports measured mass excesses for 2457 nuclei in total. This includes 2386 previously known nuclei and an extrapolation set of 71 nuclei newly measured and added only in AME2020. These additional nuclei, absent in AME2016, lie close to the nuclear driplines, where experimental information remains scarce. The different datasets employed for training and validation are summarized in (Table~\ref{supplementary_tab2}).

The dataset is organized into distinct subsets for systematic evaluation. The training set (2,028 nuclei, $\sim$85\%) and test set (358 nuclei, $\sim$15\%) are drawn from 2386 data, allowing in-domain learning and benchmarking. A new-data set of 31 nuclei measured after AME2020 provides out-of-sample validation. The all 2020 set of 2,457 nuclei enables comprehensive comparison with global nuclear mass models. Finally, the full dataset contains 2488
as described in the (Table~\ref{supplementary_tab2}).

    
    
    
    

This setup allows us to fairly compare how well the models work on both well-known nuclei and on new, less explored ones.

\section{Raw and Corrected Residuals}

\begin{figure*}[h!]
    \centering
    \includegraphics[width=0.9\textwidth, keepaspectratio]{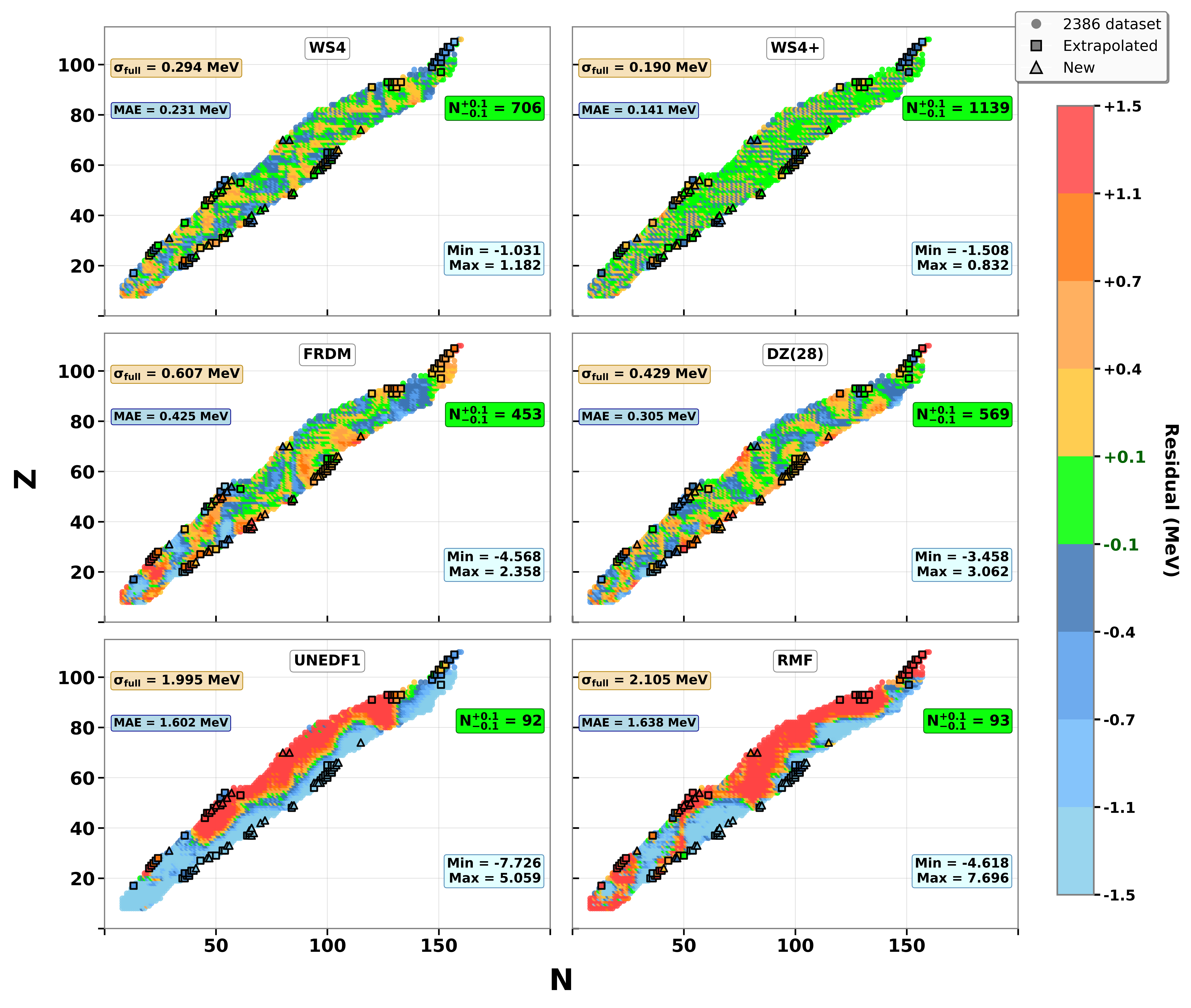}
    \caption{The residuals in nuclear mass excess ($\Delta M_0$), defined as the deviation between measured and theoretical values, are presented as functions of neutron number ($N$) and proton number ($Z$) for six nuclear mass models: WS4, WS4+, FRDM, DZ(28), UNEDF1 and RMF. The panels display the residuals for 2,386 previously known nuclei (shown by circle) and residuals for extrapolated (71) newly measured nuclei (shown by square) added only in AME2020. Newly measured 31 nuclei after AME2020 shown by triangle. A blue-to-red colour scale is used to represent the residuals: blue indicates overbinding (theoretical mass lower than experimental), red indicates underbinding, and green–yellow corresponds to values near zero. Each subplot includes the standard deviation and the number of nuclei that have values lying in the green region that is -0.1 MeV to +0.1 MeV.}
    \label{supplementary_fig1}
\end{figure*}

To study the baseline predictive performance of the selected nuclear mass models, we computed the residuals in nuclear mass excess as the difference between measured and theoretical values for each model across six different datasets as listed in (Table~\ref{supplementary_tab3}). This observation is consistent with the visual residual distribution in Figure~\ref{supplementary_fig1}, where we have displayed residuals for all the nuclei considered in various datasets across the nuclear chart. These findings collectively suggest that while some nuclear mass models offer competitive accuracy on well-characterized regions of the nuclear chart, their predictive reliability declines for extrapolated and newly measured nuclei.


\begin{table}[h!]
\centering
\caption{Comparison of root mean square errors ($\sigma$) for raw residuals across various nuclear mass models. The values of $\sigma$ are obtained for various datasets mentioned in Table~\ref{supplementary_tab2}. In the last column, we also list the values of the weights required for the averaging procedure.}
\scriptsize
\resizebox{\columnwidth}{!}{%
\begin{tabular}{|l|c|c|c|c|c|c|c|}
\hline
\multirow{2}{*}{\textbf{Models}} & 
\multicolumn{6}{c|}{$\boldsymbol{\sigma}$ (MeV)} & 
\multirow{2}{*}{$\boldsymbol{W_{0}}$} \\
\cline{2-7}
 & train & test & extpl & 2020 & new & full & \\
\hline
WS4      & 0.2895 & 0.3092 & 0.3556 & 0.2946 & 0.2892 & 0.2945 & 0.1954 \\
WS4+     & 0.1840 & 0.1735 & 0.3401 & 0.1889 & 0.2893 & 0.1905 & 0.6204 \\
FRDM     & 0.6040 & 0.5755 & 0.7864 & 0.6060 & 0.8478 & 0.6096 & 0.0564 \\
DZ(28)   & 0.4238 & 0.3972 & 0.6330 & 0.4276 & 0.5782 & 0.4298 & 0.1184 \\
UNEDF1   & 1.9962 & 1.9169 & 1.9526 & 1.9836 & 2.6935 & 1.9940 & 0.0052 \\
RMF      & 2.0756 & 2.0972 & 2.8321 & 2.1044 & 2.0880 & 2.1041 & 0.0042 \\
\hline
\end{tabular}%
}
\label{supplementary_tab3}
\end{table}

 \begin{table}[htbp]
\centering
\scriptsize
\caption{Values of RMSE ($\sigma$ in MeV) for different nuclear mass models for GBR and BAR across different subsets of data.}
\resizebox{\columnwidth}{!}{%
\begin{tabular}{|l|l|c|c|c|c|c|c|c|}
\hline
\multirow{2}{*}{\textbf{Method}} & 
\multirow{2}{*}{\textbf{Models}} & 
\multicolumn{6}{c|}{$\boldsymbol{\sigma}$ (MeV)} & 
\multirow{2}{*}{\textbf{W}} \\
\cline{3-8}
 &  & train & test & extpl & 2020 & new & full &  \\
\hline
\multirow{6}{*}{\textbf{GBR}} 
& WS4    & 0.0105 & 0.1583 & 0.2217  & 0.0719  & 0.3579 & 0.0818 & 0.1212  \\
& WS4+   & 0.0087 & 0.1216 & 0.2126  & 0.0594  & 0.2675 & 0.0661 & 0.2053  \\
& FRDM   & 0.0130 & 0.1714 & 0.2836  & 0.0821  & 0.2941 & 0.0880 & 0.1033  \\
& DZ(28) & 0.0115 & 0.1749 & 0.2709  & 0.0818  & 0.2873 & 0.0874 & 0.0993  \\
& UNEDF1 & 0.0137 & 0.2739 & 0.4394  & 0.1291  & 0.4081 & 0.1361 & 0.0405  \\
& RMF    & 0.0226 & 0.2931 & 0.5638  & 0.1487  & 0.5204 & 0.1588 & 0.0354  \\
\hline
\multirow{6}{*}{\textbf{BAR}} 
& WS4    & 0.1663 & 0.2300 & 0.2354  & 0.1793  & 0.3358 & 0.1821 & 0.0574  \\
& WS4+   & 0.0967 & 0.1165 & 0.2359  & 0.1063  & 0.2720 & 0.1099 & 0.2237  \\
& FRDM   & 0.1984 & 0.2894 & 0.4499  & 0.2248  & 0.4612 & 0.2292 & 0.0363  \\
& DZ(28) & 0.1759 & 0.2633 & 0.4438  & 0.2033  & 0.4140 & 0.2073 & 0.0438  \\
& UNEDF1 & 0.2822 & 0.4101 & 0.8044  & 0.3300  & 0.6316 & 0.3355 & 0.0180  \\
& RMF    & 0.3034 & 0.4386 & 0.7309  & 0.3456  & 0.5318 & 0.3485 & 0.0158  \\
\hline
\end{tabular}
\label{supplementary_tab4}
}
\end{table}

We corrected the residuals using two ensemble learning methods, GBR and BAR. In total, twelve sets of corrected residual were generated—six with GBR and six with BAR. The corresponding root mean square errors (RMSEs) for all different datasets are summarized in Table~\ref{supplementary_tab4}. In the last column, we have also listed the values of the weights required for the averaging procedure.
\bibliographystyle{elsarticle-num}
\vspace{-0.5cm}
\bibliography{references}